\documentclass[a4paper,1p]{elsarticle}
\pdfoutput=1
\usepackage{algorithmic}
\usepackage{tikz}
\usepackage{pgfplots}
\journal{Hal}
\usepackage[normal,small]{subfigure}
\newcommand{\lgca}{{\sc Lgca}}
\newcommand{\ca}{{\sc ca}}
\newcommand{\hpp}{{\sc Hpp}}
\newcommand{\fhp}{{\sc Fhp}}
\newcommand{\Dim}[2]{$#1\times{}#2$}
\newcommand{\Dime}[1]{\Dim{#1}{#1}}
\graphicspath{{COURBES/}}
\begin{document}
\begin{frontmatter}
\title{Lattice Gas Symmetric Cryptography}
\author{Laurent Signac}
\ead{laurent.signac@univ-poitiers.fr}
\address{Laboratoire d'Informatique et d'Automatique pour les Syst\`emes\\
\'Ecole Nationale Supérieurs d'Ingénieurs de Poitiers\\Universit\'e de Poitiers, France}
\date{\today}

\begin{abstract}
Lattice gas cellular automata ({\sc Lgca}) are particular cellular automata that
imitate the behavior of particles moving on a lattice. 
We used a particular set of \lgca{} rules, called {\sc hpp}, to
mix bits in data blocks and obtain a symmetric cryptographic
algorithm. The encryption and decryption keys are the positions of
perturbation sites on the lattice (walls). Basically, this paper
presents an original way to perform cryptographic operations, based on
cellular automata. In this paper, we show
several characteristics about our algorithm:
typical block size ($2^{2n-1}$), key-length ($2^n$), number of rounds
($2^{n+1}$). We also evaluate avalanche and strict avalanche
properties with respect to key and plain text. Finally, we highlight
the underbellies of our method and give clues to solve them.  
\end{abstract}
\begin{keyword}
cryptography, cellular automata, lattice gas, block cipher
\end{keyword}
\end{frontmatter}

\section*{Introduction}
We propose to use a particular kind of cellular
automata, called lattice gas cellular automata (\lgca), as a
symmetric cryptographic 
algorithm: Encryption and decryption are achieved by the same
algorithm and same key. One of the advantages of this algorithm is the number of
parameters that can be modified, such as the size of data blocks, the
size of keys, number of rounds... In the first section, we recall what
is symmetric block cryptography and how the issue of cellular automata
cryptography has been addressed yet.
In section 2, we introduce lattice gas cellular automata which may be new to
the cryptography community. In the third section, we detail how a
particular set of rules, called {\sc hpp} \cite{hard73} can be used to
encrypt any data. In section~4, we show how the different
algorithm parameters (number of rounds, key size...) influence the
ciphered block using gray levels images. In the last section, we check
the avalanche
and strict avalanche properties and point out the weaknesses of our
method.
Then, we conclude.

\section{Cellular automata and Cryptography}
Cryptography techniques may be divided into two categories:
symmetric-key and public-key algorithms. Using symmetric key
cryptography means that either the same key is used for encryption and
decryption or it is easy to obtain one key from another.

On the contrary, if the sender and the receiver do not use the same
key, and it is hard to compute the receiver's key from the sender's
one, we use asymmetric cryptography.

This paper deals with the first category algorithms. More precisely,
symmetric cryptography falls into stream or block ciphers. A stream
cipher encrypts one byte at a time whereas a block cipher breaks up
the message into fixed length blocks and encrypts one block at a time.
This work is about a cellular automaton based symmetric block cipher.

Connecting cryptography and cellular automata (\ca) is not new. But every
cryptography topic do not have the same relationship with cellular
automata. First of all, during the 80's, Wolfram~\cite{wolfram86}
showed that \ca{} were good pseudo random number generators. As random number
generators have a very important role in stream cipher algorithms, \ca{}
based stream cipher algorithms have been widely
studied~\cite{Seredynski03,tomassini00,tomassini01,wolfram85,Seredynski2004753}.
The common principle of these works is to use the initial \ca{} state
as a seed and to run
the \ca{} in order to generate a random sequence. This sequence is then
used in the stream cipher algorithm (Vernam cipher for instance).

On the opposite, \ca{} based public key cryptography literature is not
abundant. See for instance~\cite{guan87,kari92,gutowitz93}.

Between the two opposites, several papers tackle the issue of
symmetric block encryption. Most of them use a key built on the \ca{}
rules~\cite{kari92,seredynski04}.

From this point of view, our approach is quite different.
Indeed, in our algorithm, keys are
encoded in the \ca{} state rather than in the \ca{} rules.
This is due to the particular kind of \ca{} we
used: Lattice Gas Cellular Automata. 

Nevertheless, we recently found a very similar approach
in~\cite{chopard06} (the present work was initiated a (very) long
time ago...). In paper~\cite{chopard06}, the authors also propose a block
cipher symmetric cryptography based on \lgca. Contrary to us, they do not 
focus on a particular \textsc{lgca} but tackle the issue in a more general
way. The secret key in~\cite{chopard06} is \textsc{xor}ed each step
with the lattice state (see section~2.1) whereas we added a
particular step called \textit{reflection} in our algorithm to process
the key (to a certain extend, the two approaches are similar, although our is more
\textit{visual}, but less general).
Besides, some remarks are similar : number of rounds, propagation of
error. Finally, differential cryptography issue is tackled
in~\cite{chopard06} whereas we focused on avalanche and strict
avalanche properties.
 
\section{Lattice Gas Cellular Automata}

The world of \ca{}~\cite{ulam52,neum66} gave birth to particular
simulation methods called Lattice Gas Cellular Automata
(\lgca)~\cite{wolf00}.
{\sc Lgca} are designed to mimic 
the moves of many particles on a grid. The first \lgca{} has
been proposed in 1973 by Hardy, Pazzis and Pomeau \cite{hard73}.
It is this particular \lgca{} called \hpp{} that we use in this
paper. 

\subsection{{\sc Hpp}}
{\sc Hpp} is a two-dimensional cellular automaton living on a square
lattice. The Von Neumann neighborhood is used: each cell has four
neighbors.
Each cell may be occupied by up to four particles: the east, the
north, the west and the south particle. 

\begin{figure}[htbp]
\begin{center}
\includegraphics[page=1,scale=0.9]{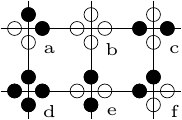}
\end{center}
\caption{A sample lattice with 6 cells}
\label{fig:gridsample}
\end{figure}

Figure~\ref{fig:gridsample} shows a lattice made of 6 cells. The upper-left corner
cell is occupied by the north and the east particles. Its right
neighbor is empty and its bottom neighbor is occupied by the four
particles.

It is convenient to represent the state of a cell by a four bits
number: the most significant bit will be the east
particle, followed by the south, west and north particles. 

The upper line of Figure~\ref{fig:gridsample} can be represented
as: 1001 0000 1010 ($9\,0\,A$ in hexa).

Evolution of particles in \hpp{} is a two steps  algorithm. First,
particles collide in a cell, then particles propagate.

Among the 16 possible cell configurations, particles do collide in only two
particular configurations: 1010 and 0101. 
For instance, on Figure~\ref{fig:gridsample},
particles in cell $c$ and $e$ will collide. No collision will occur in
other cells.

A west-east collision will give a north-south configuration, and a
north-south collision will give a west-east configuration.

\begin{figure}[htbp]
\begin{center}
\includegraphics[page=2,scale=0.9]{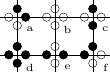}
\end{center}
\caption{The sample lattice (fig. \ref{fig:gridsample}) after the collision step}
\label{fig:aftercollision}
\end{figure}

After the collision step, the lattice of Figure~\ref{fig:gridsample}
will turn into the lattice of Figure~\ref{fig:aftercollision}.

The propagation step follows. During this step, every particle will move. The north
particle of a given cell will move up (and remain a north
particle). The west particle will move
left (and remain a west particle) {\it etc}.

\begin{figure}[htbp]
\begin{center}
\includegraphics[page=3,scale=0.9]{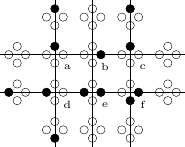}
\end{center}
\caption{The sample lattice (fig.~\ref{fig:gridsample}) after the
  collision (fig.~\ref{fig:aftercollision} and the propagation step}
\label{fig:afterpropagation}
\end{figure}

After the propagation step, the lattice of Figure~\ref{fig:aftercollision}
will turn into the lattice of Figure~\ref{fig:afterpropagation}. 

We have seen how particles move on a 2-dimensional infinite
lattice. The \hpp{} rules may also be used on a finite
lattice by connecting the upper and lower bounds as well as the right
and left bounds of the lattice (torus topology).

\begin{figure}[htbp]
\begin{center}
\includegraphics[page=4,scale=1]{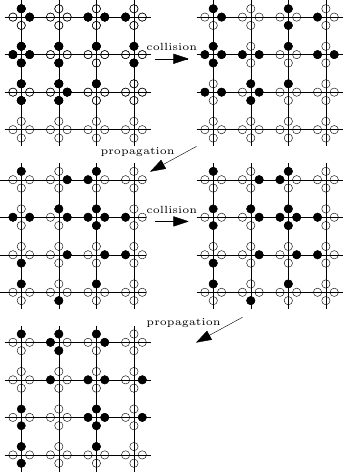}
\end{center}
\caption{Particles evolving on a toric lattice}
\label{fig:evolve}
\end{figure}

Figure~\ref{fig:evolve} shows the first evolution rounds of a \Dim{4}{4} toric
lattice. As mentioned above, the set of cells may be represented as
set of \Dim{16}{4} bits {\it i.e.} 8 bytes. The top-left corner cell will be the
most significant 4 bits of the first byte. The cell on its right will
be the less significant 4 bits of the first byte and so on. 
In the first step, the top left corner site contains west and north
particles ($1001 \rightarrow 9$). Its right neighbor contains no
particle ($0$). The third site of the top line contains east and west
particle ($1001 \rightarrow A$). Le last site only contains the west
particle ($0010\rightarrow 2$).

Here are the 8 bytes representing the initial state, and the 2 states obtained
after the propagation step (hexadecimal numbers):
\begin{itemize}
\item {\tt 90 A2 F5 15 5D 10 00 00}
\item {\tt 18 30 A9 F2 48 82 14 10}
\item {\tt 17 90 02 A8 50 F8 50 10}
\end{itemize}

Now, we know how to make a set of bytes evolve. Let us add perturbation
sites.

\subsection{Obstacles Reflections}
We now introduce special cells known as 'walls' or 'obstacles', and a step called
'reflection'. After the collision and propagation steps, direction of particles
that live on a wall cell are inverted, so that the north and south, as
well as east and west are swapped. 

\begin{figure}[htbp]
\begin{center}
\includegraphics[page=5,scale=1.2]{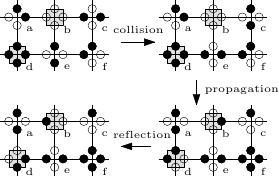}
\end{center}
\caption{Collision, propagation and reflection step on two walls}
\label{fig:reflection}
\end{figure}

Figure~\ref{fig:reflection} shows an example of a $3\times2$ lattice with two
'walls' on cells~$b$ and~$d$.

\section{HPP Symmetric Cryptography}
In this section, we show how \hpp{} rules could be used as a symmetric
cryptographic algorithm. We first recall that \hpp{} simulations are
reversible. Then we detail the steps of the cryptographic algorithm.
Finally, we propose to use obstacles positions as a secret key.

\subsection{Time Reversibility}
The most important property of \hpp{} rules we used to design the cryptographic algorithm is time
reversibility. In other words, we used the fact that, during
particles evolutions, no information is lost, and the initial
configuration can be computed from any subsequent configuration.

Even though we focused on \hpp{} rules, any \lgca{} set of rules could make the
job. The condition is that the set of rules is time reversible. This is
not the case for \fhp{} rules~\cite{fris86}, on an hexagonal lattice:  the
collision step, for physical realism reasons (chirality), is
probabilistic. As we are not interested in physical realism, \fhp{}
rules could be modified to be time reversible.

\subsection{Cryptographic Algorithm}
The way we used \hpp{} as a cryptographic algorithm may now be
clear:
\begin{enumerate}
\item Choose a lattice size. It may be convenient to choose a power of
  two. For instance \Dim{256}{256}.
\item Segment raw data into blocks to populate the lattice. For a \Dime{256} grid,
  we need 32768  bytes data blocks, as every cell is
  populated with 4 bits.
\item Choose a secret key, and use this secret key to compute the walls
  positions. The way how a secret key is chosen and how it is used will be
  tackled in section~3.3.
\item Choose $n$, the number of rounds that will be computed. 
\item For each clear data block:
\begin{itemize}
\item Apply $n$ times: collision, propagation, reflection.
\item Reverse time (to get an involutive algorithm).
\item Get the encrypted data block.
\end{itemize}
\end{enumerate}

Decryption is achieved by applying exactly the same
algorithm to encrypted data blocks, provided each parameter is the
same:
\begin{itemize}
\item block size
\item secret key
\item number of rounds
\end{itemize}

\subsection{Secret Key}
The secret key is the position of the walls on the grid.
If the lattice size is a power of two, say \Dime{2^n}, the coordinates
of a cell may be represented with $2\times n$ bits. As a consequence,
for a \Dime{2^n} lattice, and $K$ walls, the size of the key must
be $2nK$ bits. However, the number of {\it real} keys is not
$2^{2nK}$, as exchanging certain walls positions might give the same key, and several
walls may be in the same cell. The number of walls configurations is
the number of multisets of cardinality $K$ of elements taken from a set
of cardinality $2^{2n}$: 
$$\left(\!\!\left(%
\begin{array}{c}2^{2n}\\K\end{array}%
\right)\!\!\right)=C_{K}^{2^{2n}+K-1}=\frac{(2^{2n}+K-1)!}{K!(2^{2n}-1)!}$$

The number of potential secret keys is quite important. If
we use a \Dime{256} lattice ($n=8$), and 256 walls ($K=256$), the number of
different walls configurations  is 
$C_{256}^{256^2+256-1}\approx 2\times 10^{726}$. 
For a small \Dime{64} lattice ($n=6$, 2048 bytes data blocks) and 32 walls
(K=32), the number of available keys is about $1.6\times 10^{80}$.

\section{Experimental Results : LGCA point of view}
In this section, we show how to choose a key and the number of
rounds. We also point out the problem of invariants which may be a
problem in cryptographic applications whereas it is a desirable
behavior in {\sc lgca}.

In this section, we used pictures as the properties we want to show
can be well understood by thinking of our algorithm as a lattice-gas
cellular automaton. 
Nevertheless, one must keep in mind that our block cipher is not
particularly designed to encrypt pictures, but any data block.

Each cell of the automaton may be
represented by a 4 bits number. We use this number as a gray
level. Thus, a $2^n\times2^n$ lattice populated with particles is
turned into a 16 gray levels image of size $2^n\times2^n$.

\begin{figure}[htbp]
\begin{center}
\subfigure[Eiffel tower\label{fig:eiffel}]{\includegraphics[width=.45\textwidth]{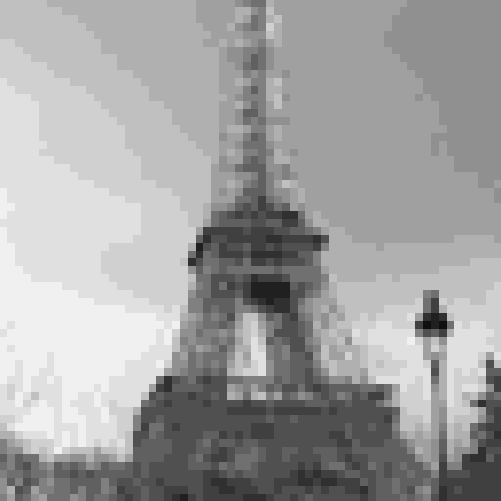}}
\subfigure[A squirrel\label{fig:squirell}]{\includegraphics[width=.45\textwidth]{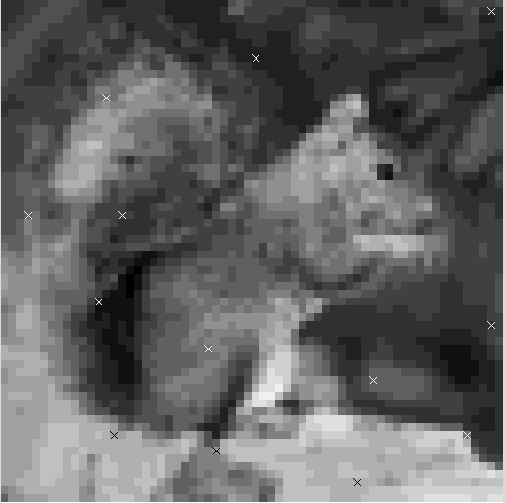}}
\end{center}
\caption{Images used in our tests}
\label{fig:imagetest}
\end{figure}

In the following, we used two \Dime{64} images (Figure~\ref{fig:imagetest}).

\subsection{Number of Rounds}
\begin{figure}[htbp]
\begin{center}
\includegraphics[width=.45\textwidth]{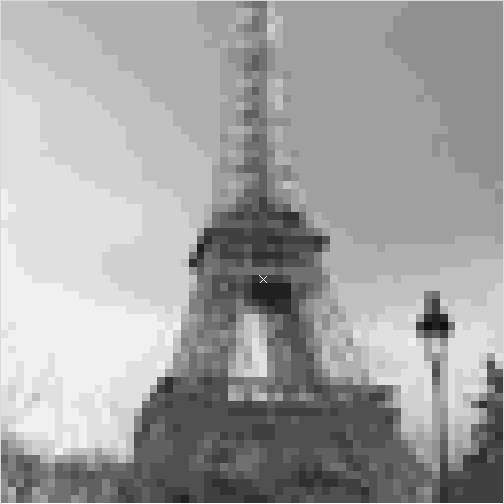}
\end{center}
\caption{Clear picture, with a central wall (small white cross)}
\label{fig:eiffel1}
\end{figure}

\begin{figure}[htbp]
\begin{center}
\subfigure[Encrypted picture]{\includegraphics[width=.45\textwidth]{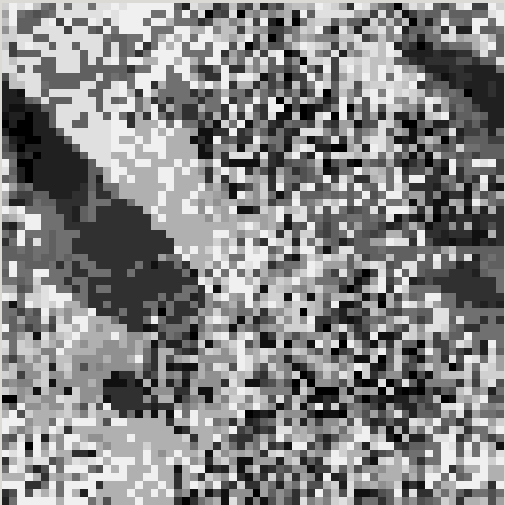}}
\subfigure[Decrypted picture, without the central wall]{\includegraphics[width=.45\textwidth]{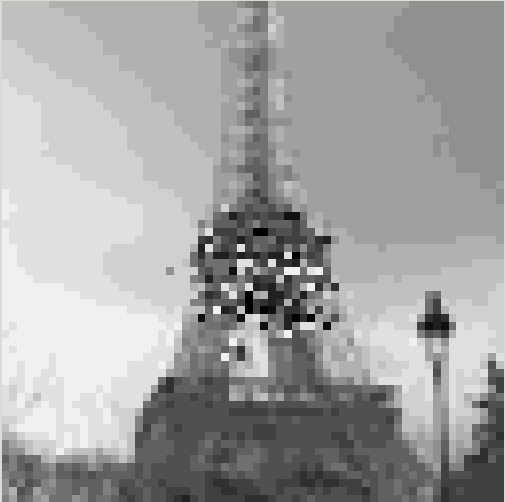}}
\end{center}
\caption{Encrypted picture, and the decrypted picture obtained by
  omitting the central wall (16 rounds)}
\label{fig:eiffel1_16}
\end{figure}

\begin{figure}[htbp]
\begin{center}
\subfigure[Encrypted picture]{\includegraphics[width=.45\textwidth]{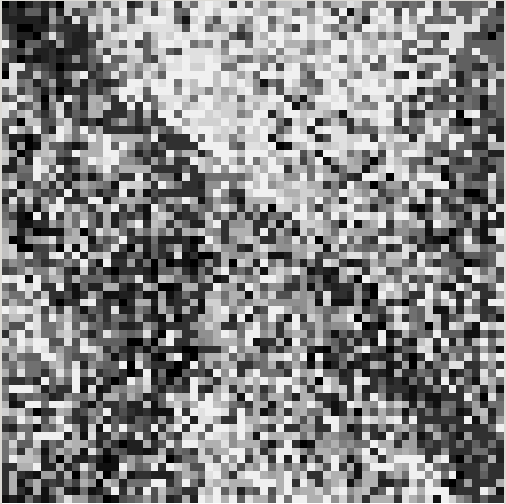}}
\subfigure[Decrypted picture, without the central wall]{\includegraphics[width=.45\textwidth]{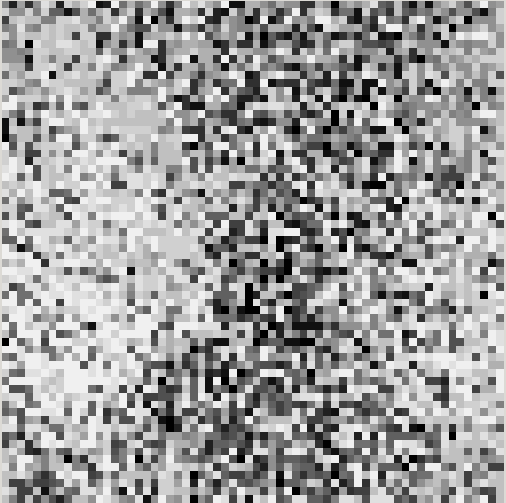}}
\end{center}
\caption{Encrypted picture, and the decrypted picture obtained by
  omitting the central wall (128 rounds)}
\label{fig:eiffel1_128}
\end{figure}

First, we encrypt our image, with only one wall (look at the (very) small
cross) in the center
(Figure~\ref{fig:eiffel1}). Then, we remove the wall, and try to
decrypt the picture. We first make this experiment with a 16 rounds
algorithm~(Figure~\ref{fig:eiffel1_16}), and then with a 128 rounds
algorithm (Figure~\ref{fig:eiffel1_128}).

We see that if the number of rounds is not large enough, a part of the image can be
decrypted without the key. This is an evidence that a wall can not
influence cells that are too far. More precisely, by computing $N$
rounds, cells that are outside the discus of radius $N$ (we consider the
Hamming distance) and centered on the wall will not be affected by the
wall.  Cells that are inside the discus may or may not be affected:
it depends on the particles positions, that is not known {\it a
  priori}.

The consequence is that we must choose a large enough value for $N$ so
that every cell is affected by every walls, whatever the position of
the walls.

Let us choose a number of rounds such that
{\em every} cell can be affected by every wall. For a $2^n\times 2^n$
lattice, the maximal Hamming distance is $2^{n}$ (keep in mind the
lattice is on a torus). By choosing
$N=2^{n}$, we avoid the problem of key-unaffected bits. 
Unfortunately, computing $2^{n}$ rounds may be time consuming. 

One may think that, because of the large number of walls, it is
possible to reduce the number of rounds. It would not be a good idea.

\begin{figure}[htbp]
\begin{center}
\subfigure[Original image. No wall next to the head\label{fig:squi}]%
{\includegraphics[width=.45\textwidth]{IMG/squi.png}}
\subfigure[Encrypted image with 24 rounds\label{fig:squienc24}]%
{\includegraphics[width=.45\textwidth]{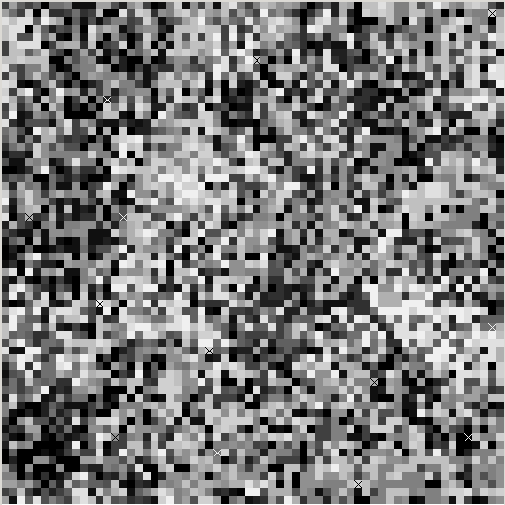}}
\subfigure[Decrypted image with a wrong key\label{fig:squidec24}]%
{\includegraphics[width=.45\textwidth]{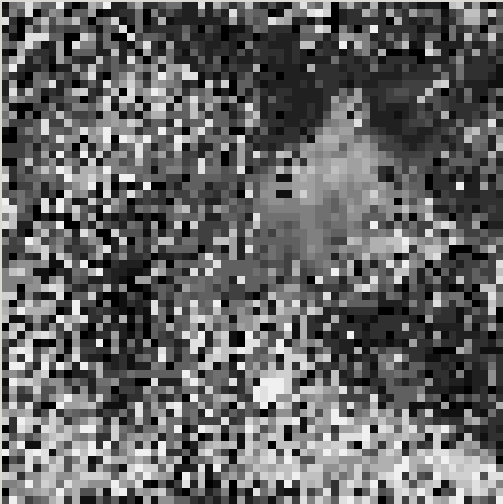}}
\end{center}
\caption{A part of the image can be decrypted if the number of rounds
  is too low: the squirrel's head is visible}
\end{figure}

Here is another example: the squirrel of Figure~\ref{fig:squi} is
encrypted with a 13 walls key. Unfortunately, the keys are not well
spread and there is no wall next to the squirrel's head.
Using a number of rounds too low (say 24), we obtain the encrypted
picture of Figure~\ref{fig:squienc24}. It is then possible to decrypt
a part of the image (squirrel's head) by applying the algorithm
with a key that does not contain a wall next to the head.
We then obtain the picture of Figure~\ref{fig:squidec24}.

This would be impossible with a higher number of rounds.

The only way to avoid (to a certain extend) the decryption of parts of an image by only knowing
parts of the key, is to choose a number of rounds such
that each wall influences every cell.

\begin{figure}[htbp]
\begin{center}
\subfigure[Encrypted image with 32 walls and 64 rounds\label{fig:squienc64_k32}]%
{\includegraphics[width=.45\textwidth]{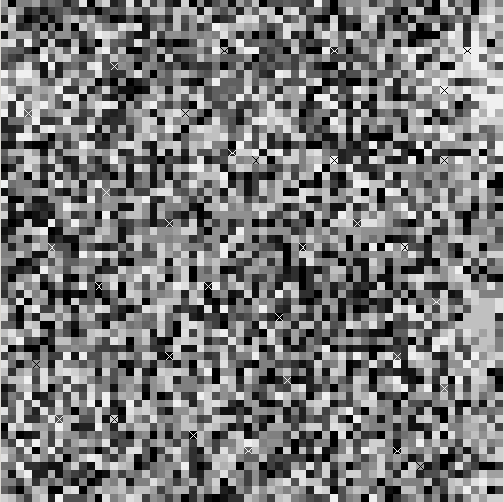}}
\subfigure[Decrypted image with very slightly different key\label{fig:squidec64_k32}]%
{\includegraphics[width=.45\textwidth]{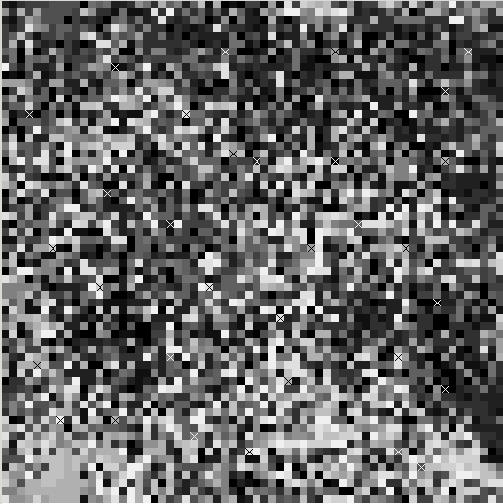}}
\end{center}
\caption{With a large key and a large number of rounds, the algorithm
  is very sensible to key variations}
\end{figure}

By using a key of about  $2^{n-1}$ walls ($2^n\times2^n$ is the size of
the image), and a number of rounds of $2^n$, the algorithm would be very
sensible to key variations. Figure~\ref{fig:squienc64_k32} shows the \Dime{64}
squirrel encrypted with a 32 walls key and 64 rounds.

This encrypted image is decrypted by a key that differs only on one
wall, that is shifted right one cell. The 'decrypted' picture is shown
on Figure~\ref{fig:squidec64_k32}

In section~\ref{sec:avalanche}, we will see that, for a \Dime{64}
lattice (2048 bytes), a good number of rounds is in fact 128.

\subsection{Algorithms Invariants}
{\sc Hpp}, as well as most \lgca{} rules, have been designed to
simulate physical systems. One characteristic is that \lgca{} obey
physical laws such as mass conservation, momentum conservation...

For our particular application, this is a great disadvantage. For
\hpp{} model, mass and momentum are the same, because each particle
has the same velocity. More informations about real or spurious
invariants in \lgca{} can be found in~\cite{wolf00}. 

For our application, it means that the number of bits equal to 1
remains constant in a data block. This is very annoying, especially if
the number of 1's (or 0's) is very low. 

On our gray levels pictures, it means that a black picture will remain
black, whatever the number of walls, their positions, and the number of
rounds... 

We conclude that the algorithm must not be used on data blocks that
contain redundant informations.

Let us consider the issue in this way: the number of bits equal to 1 in
the clear block can be deduced from the number of bits equal to 1 in
the encrypted block. How to make this information useless? 
We could for instance only consider data blocks where the number of 1's
is {\em always} about half the number of bits. If the algorithm is
only applied on such blocks, the knowledge of the number of 1's becomes
useless. 

Consequently, we need a preprocessing step that turns any data
blocks into data blocks containing as many 1's as 0's. This job may
be done with almost any compression algorithm, that theoretically
produce random-like data ({\sc Gnu/}gzip utility
(Lempel-Ziv algorithm) for instance).

Using compression mainly solves the mass invariance of
\lgca. Furthermore it speeds up encryption.

\section{Experimental Results : cryptography point of view}
\label{sec:avalanche}
\subsection{Avalanche property}
Any encryption algorithm should satisfy the following property:
a small change in the key or in the plain text should result in a great
modification of the ciphered text. More precisely, if we invert one
bit in the plain text or in the key, we expect that nearly half
ciphered text bits are inverted. This property, called \emph{avalanche
property} has been introduced in~\cite{feistel73}. 

To show the avalanche property, we
encrypt 2048 bytes blocks
($64\times64$ lattice). The key is 56 bytes long. 

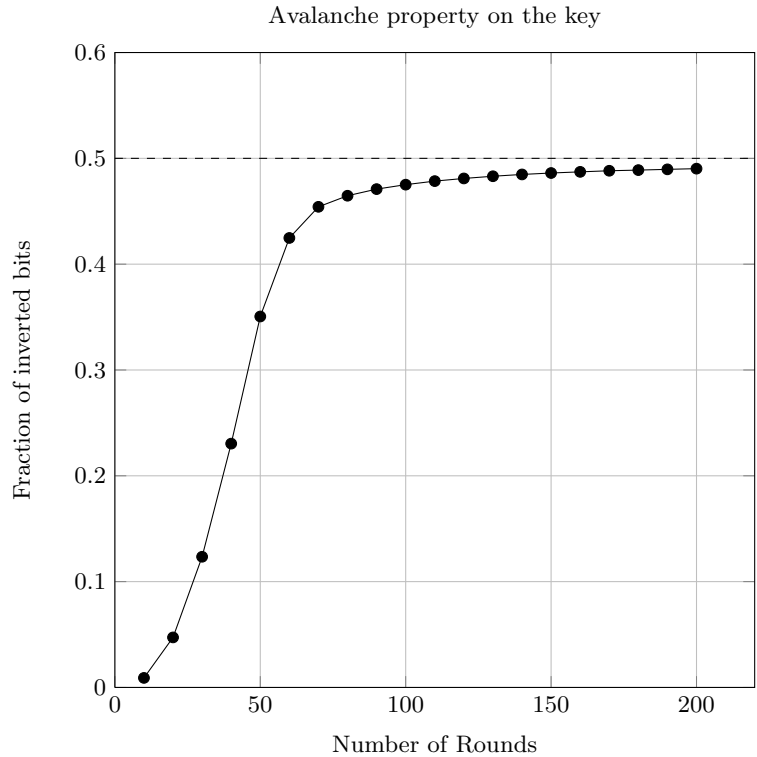
\begin{figure}[htbp]
\begin{center}
\small
\begin{tikzpicture}
\begin{axis}[height=10cm, width=10cm,grid=major,
  xmin=0,xmax=220,ymin=0,ymax=0.6,
  xlabel=Number of Rounds,
  ylabel=Fraction of inverted bits,
  title=Avalanche property on the key]
\addplot[mark=none,draw=black,dashed] coordinates{(0,0.5) (220,0.5)};
\addplot[mark=*,draw=black] coordinates {(10, 0.00900859832763672) (20, 0.047272872924804685) (30, 0.1234932581583659) (40, 0.23040536244710283) (50, 0.35055243174235023) (60, 0.4247672398885092) (70, 0.4541948954264323) (80, 0.46466941833496095) (90, 0.47092641194661466) (100, 0.4751277923583984) (110, 0.47848974863688143) (120, 0.4810029983520508) (130, 0.48308353424072265) (140, 0.4847869873046875) (150, 0.4861367543538412) (160, 0.4872335433959961) (170, 0.48827152252197265) (180, 0.4888940175374349) (190, 0.4896087646484375) (200, 0.49025726318359375)};
\end{axis}
\end{tikzpicture}
\end{center}
\caption{Avalanche property if we change one bit of the key\label{fig:avalanche1}}
\end{figure}

We used the following
algorithm to produce the Figure~\ref{fig:avalanche1} that shows the percentage of
bits that have been inverted as a function of the number of
rounds\footnote{This algorithm has been executed 5 times,
  Figure~\ref{fig:avalanche1} represents the average results}:
\begin{algorithmic}
\STATE $t\gets$ 2048 random bytes
\STATE $k\gets$ 48 random bytes
\FOR{$r=10:10:200$}
  \STATE $p\gets$ 0
  \STATE $c_{ref}\gets$ encrypt($t,k$) with $r$ rounds
  \FOR{$i=0$ \TO $383$}
      \STATE $k_2\gets$ $k$ with bit $i$ inverted
      \STATE $c_2\gets$ encrypt($t,k_2$) with $r$ rounds
      \STATE $p\gets p + $fraction of inverted bits between $c_{ref}$ and $c_2$
  \ENDFOR
  \STATE $p\gets p/384$
  \STATE plot the point $(r,p)$
\ENDFOR
\end{algorithmic}

Figure~\ref{fig:avalanche1} shows that avalanche property is satisfied
with respect to the key.

In order to check the avalanche property with respect to the plain
text bits, we used a similar algorithm, executed 20 times:
\begin{algorithmic}
\STATE $t\gets$ 128 random bytes
\STATE $k\gets$ 8 random bytes
\FOR{$r=10:2:200$}
  \STATE $p\gets$ 0
  \STATE $c_{ref}\gets$ encrypt($t,k$) with $r$ rounds
  \FOR{$i=0$ \TO $1023$}
      \STATE $t_2\gets$ $t$ with bit $i$ inverted
      \STATE $c_2\gets$ encrypt($t_2,k$) with $r$ rounds
      \STATE $p\gets p + $fraction of inverted bits between $c_{ref}$ and $c_2$
  \ENDFOR
  \STATE $p\gets p/1024$
  \STATE plot the point $(r,p)$
\ENDFOR
\end{algorithmic}

\begin{figure}[htbp]
\begin{center}
\small

\begin{tikzpicture}
\begin{axis}[height=10cm, width=10cm,grid=major,
  xmin=0,xmax=220,ymin=0,ymax=0.6,
  extra y ticks={0.25},
  xlabel=Number of Rounds,
  ylabel=Fraction of inverted bits,
  title=Avalanche property on clear text]
\addplot[mark=none,draw=black,dashed] coordinates{(0,0.25) (220,0.25)};
\addplot[mark=*,draw=black] coordinates {(10, 0.03528814315795899) (12, 0.054888343811035155) (14, 0.07987384796142578) (16, 0.10721778869628906) (18, 0.13469429016113282) (20, 0.15879421234130858) (22, 0.17791051864624025) (24, 0.19280357360839845) (26, 0.20384702682495118) (28, 0.21123552322387695) (30, 0.21742191314697265) (32, 0.2219487190246582) (34, 0.2252516746520996) (36, 0.2277052879333496) (38, 0.23001594543457032) (40, 0.23150854110717772) (42, 0.2328359603881836) (44, 0.23400840759277344) (46, 0.23475046157836915) (48, 0.23546237945556642) (50, 0.23604974746704102) (52, 0.2365407943725586) (54, 0.23701019287109376) (56, 0.23726720809936525) (58, 0.23750181198120118) (60, 0.23767824172973634) (62, 0.23803319931030273) (64, 0.2383397102355957) (66, 0.23857154846191406) (68, 0.23861026763916016) (70, 0.2386547088623047) (72, 0.2388005256652832) (74, 0.23893289566040038) (76, 0.2390645980834961) (78, 0.23916234970092773) (80, 0.23927602767944336) (82, 0.23937015533447265) (84, 0.23944368362426757) (86, 0.23944244384765626) (88, 0.23940448760986327) (90, 0.2395451545715332) (92, 0.2395599365234375) (94, 0.23964195251464843) (96, 0.2396866798400879) (98, 0.23980703353881835) (100, 0.23980426788330078) (102, 0.2398059844970703) (104, 0.23981771469116211) (106, 0.23983545303344728) (108, 0.23976726531982423) (110, 0.23981618881225586) (112, 0.2398599624633789) (114, 0.23988304138183594) (116, 0.23990917205810547) (118, 0.23984956741333008) (120, 0.23986434936523438) (122, 0.2398373603820801) (124, 0.23987321853637694) (126, 0.2398448944091797) (128, 0.23987016677856446) (130, 0.2399895668029785) (132, 0.23990154266357422) (134, 0.24003467559814454) (136, 0.23991880416870118) (138, 0.2398993492126465) (140, 0.23993864059448242) (142, 0.23998947143554689) (144, 0.24003124237060547) (146, 0.23993368148803712) (148, 0.23987836837768556) (150, 0.23991289138793945) (152, 0.2398625373840332) (154, 0.23987274169921874) (156, 0.23997325897216798) (158, 0.24007186889648438) (160, 0.2399068832397461) (162, 0.23986902236938476) (164, 0.24001150131225585) (166, 0.23994092941284179) (168, 0.2399308204650879) (170, 0.23994903564453124) (172, 0.23979625701904297) (174, 0.23989057540893555) (176, 0.23995494842529297) (178, 0.23997840881347657) (180, 0.23996238708496093) (182, 0.2399600028991699) (184, 0.2400543212890625) (186, 0.24000873565673828) (188, 0.24005661010742188) (190, 0.24011802673339844) (192, 0.2400665283203125) (194, 0.24007711410522461) (196, 0.24008531570434571) (198, 0.24007177352905273) (200, 0.24015283584594727) (202, 0.24021215438842775) (204, 0.24022626876831055) (206, 0.24011697769165039) (208, 0.24022855758666992)};
\end{axis}
\end{tikzpicture}
\end{center}
\caption{Avalanche property if we change one bit of the plain text\label{fig:avalanche2}}
\end{figure}
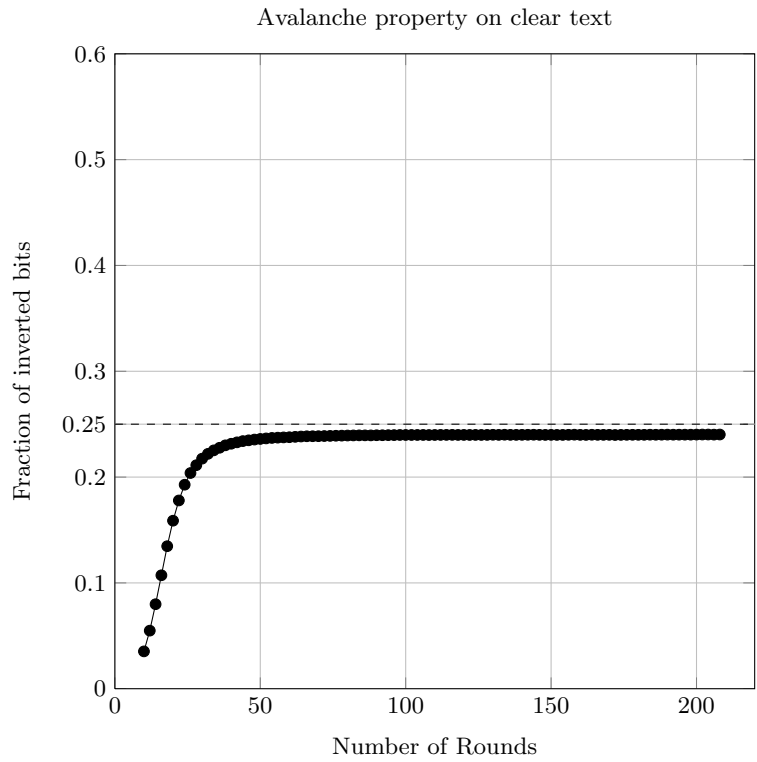

Figure~\ref{fig:avalanche2} shows the surprising results. Only 25
percent of output bits have been changed. This annoying phenomenon is explained
in the following subsection.

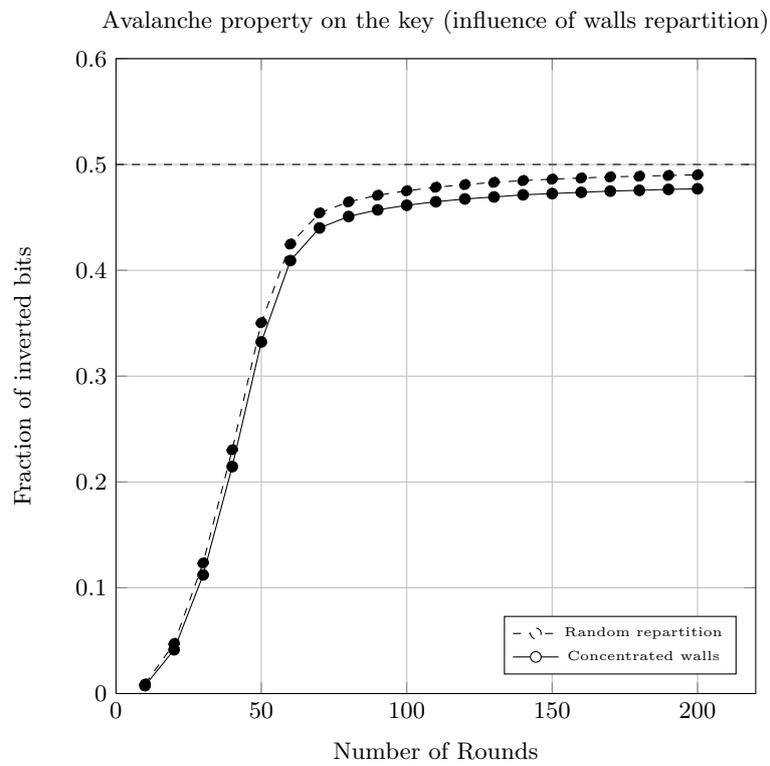
\begin{figure}[htbp]
\begin{center}
\small
\begin{tikzpicture}
\begin{axis}[height=10cm, width=10cm,grid=major,
  xmin=0,xmax=220,ymin=0,ymax=0.6,
  xlabel=Number of Rounds,
  ylabel=Fraction of inverted bits,
  title=Avalanche property on the key (influence of walls repartition),
  legend entries={\tiny Random repartition,\tiny Concentrated walls},
  legend style={at={(0.97,0.03)},anchor=south east}]

\draw [dashed] (axis cs:0,0.5) -- (axis cs:220,0.5);
\addplot[mark=*,draw=black,dashed] coordinates {(10, 0.00900859832763672) (20, 0.047272872924804685) (30, 0.1234932581583659) (40, 0.23040536244710283) (50, 0.35055243174235023) (60, 0.4247672398885092) (70, 0.4541948954264323) (80, 0.46466941833496095) (90, 0.47092641194661466) (100, 0.4751277923583984) (110, 0.47848974863688143) (120, 0.4810029983520508) (130, 0.48308353424072265) (140, 0.4847869873046875) (150, 0.4861367543538412) (160, 0.4872335433959961) (170, 0.48827152252197265) (180, 0.4888940175374349) (190, 0.4896087646484375) (200, 0.49025726318359375)};
\addplot[mark=*,draw=black] coordinates {
(10, 0.007503237985801206) (20, 0.04139829384536191) (30, 0.11212876850718327) (40, 0.21448306356408292) (50, 0.33226932679607946) (60, 0.4091632232272892) (70, 0.44002514529002984) (80, 0.45082504989442435) (90, 0.4571277519659446) (100, 0.4614661533925273) (110, 0.4647643029272973) (120, 0.4674353017234873) (130, 0.469291252329984) (140, 0.4713189699886445) (150, 0.47250857670867885) (160, 0.47367191494878647) (170, 0.4748127693604024) (180, 0.47559117857860855) (190, 0.4764488936260337) (200, 0.47703482841161293)
};
\end{axis}
\end{tikzpicture}
\end{center}
\caption{Avalanche property: comparison between a random repartition
  of walls and a configuration with walls concentrated in a small surface\label{fig:petitescles}}
\end{figure}

One can think that the distribution of walls is of great
importance. We checked the behaviour of the avalanche property for key
walls restricted to a small area of the grid (a $8\times8$ subgrid,
the whole grid being a $64\times64$ grid). Figure
\ref{fig:petitescles} gives the results compared to what we obtained
on Figure~\ref{fig:avalanche1}. 
We can see that increasing the round numbers counterbalances the
non-random repartition of walls.

\subsection{Strict avalanche properties}
Another important property is that, for every input value (key and
plain text),  a single bit change must affect the
whole ciphered text: every output bit must depend upon all input
bits.
This property, known as completeness has been combined
with the avalanche property: each output bit should change with a
probability of one half whenever a single input bit is
complemented. This property is known as \emph{strict avalanche
property}~\cite{webster85}. 

Let us look closer at this property. We have to check it for all input
bits. First we are going to check it for the key bits, and then for
the plain text bits.

The following experiments were made with small data blocks (128
bytes$\rightarrow$ 1024 bits), and
small keys (8 bytes). The number of rounds was 64, and the following
algorithm has been executed for $N=1\,000$.

\begin{algorithmic}
\STATE $v\gets {0,0,0...,0,0,0}$
\FOR{$n=1$ \TO $N$}
  \STATE $p\gets$ 128 bytes random plain text
  \STATE $k\gets$ 8 bytes random key
  \STATE $c\gets$ encrypt($p,k$)
  \FOR{$i=0$ \TO 63}
    \STATE $k_2\gets$ $k$ with bit $i$ inverted
    \STATE $c_2\gets$ encrypt($p,k_2$)
    \STATE $v\gets v + c \mbox{~xor~} c_2$ \COMMENT{$v,c,c_2$ are arrays}
  \ENDFOR
\ENDFOR
\STATE $v\gets v/(N\times64)$
\end{algorithmic}

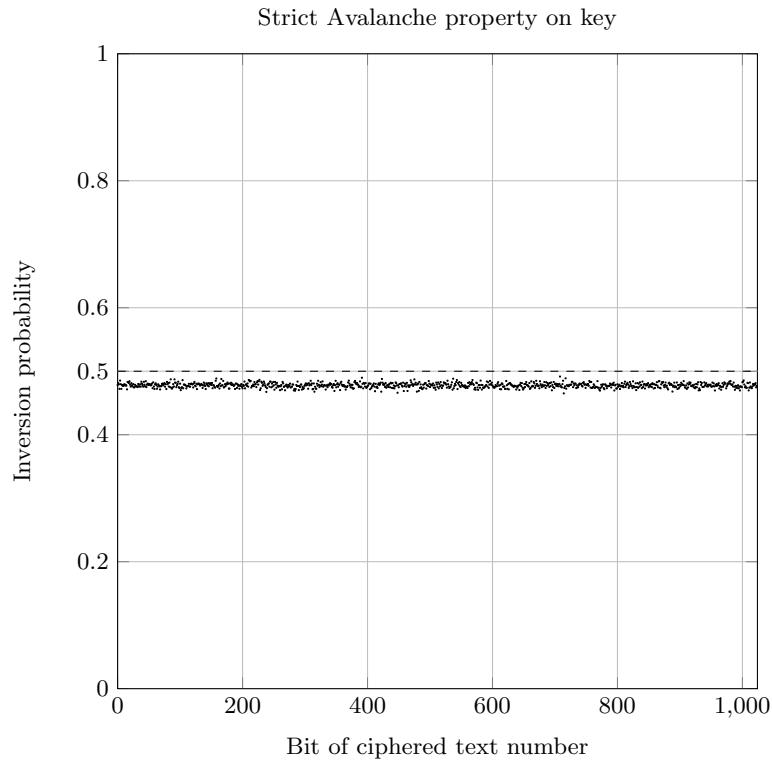
\begin{figure}[htbp]
\begin{center}
\small
\begin{tikzpicture}
\begin{axis}[height=10cm, width=10cm,grid=major,
  xmin=0,xmax=1024,ymin=0,ymax=1.0,
  extra y ticks={0.5},
  xlabel=Bit of ciphered text number,
  ylabel=Inversion probability,
  title=Strict Avalanche property on key]
\addplot[mark=none,draw=black,dashed] coordinates{(0,0.5) (1023,0.5)};
\addplot[only marks,mark=*,mark options={xscale=0.1,yscale=0.1}] file {data/strictavalanche_key.data};
\end{axis}
\end{tikzpicture}
\end{center}
\caption{Influence of changing a bit in the key :
  each output bit has a probability of about $0.47$ to be inverted
\label{fig:strict_lowkey}}
\end{figure}

The results, given on Figure~\ref{fig:strict_lowkey} show that
if we change one bit of the key, the probability for \emph{every}
ciphered bit to be inverted is about 0.47.
 
Now, let us take a look at the influence of plain text bits. The following algorithm
has been used ($N=1\,000$):
\begin{algorithmic}
\STATE $v\gets {0,0,0...,0,0,0}$
\FOR{$n=1$ \TO $N$}
  \STATE $p\gets$ 128 bytes random plain text
  \STATE $k\gets$ 8 bytes random key
  \STATE $c\gets$ encrypt($p,k$)
  \FOR{$i=0$ \TO 1023}
    \STATE $p_2\gets$ $p$ with bit $i$ inverted
    \STATE $c_2\gets$ encrypt($p_2,k$)
    \STATE $v\gets v + c \mbox{~xor~} c_2$ \COMMENT{$v,c,c_2$ are arrays}
  \ENDFOR
\ENDFOR
\STATE $v\gets v/(N\times1024)$
\end{algorithmic}

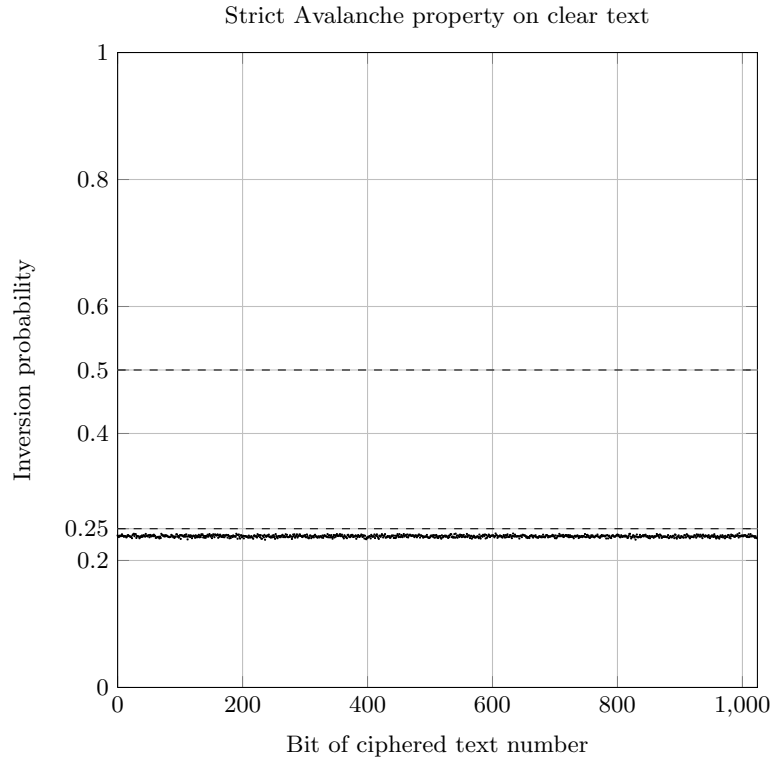
\begin{figure}[htbp]
\begin{center}
\small
\begin{tikzpicture}
\begin{axis}[height=10cm, width=10cm,grid=major,
  xmin=0,xmax=1024,ymin=0,ymax=1.0,
  extra y ticks={0.25,0.5},
  xlabel=Bit of ciphered text number,
  ylabel=Inversion probability,
  title=Strict Avalanche property on clear text]
\addplot[mark=none,draw=black,dashed] coordinates{(0,0.5) (1023,0.5)};
\addplot[mark=none,draw=black,dashed] coordinates{(0,0.25) (1023,0.25)};
\addplot[only marks,mark=*,mark options={xscale=0.1,yscale=0.1}] file {data/strictavalanche_text.data};
\end{axis}
\end{tikzpicture}
\end{center}
\caption{Influence of a bit inversion on plain text
\label{fig:strict_text}}
\end{figure}

Figure~\ref{fig:strict_text} shows the result. Again, for a bit
inversion in the clear text, the probability for a bit in the ciphered
text to be inverted is 25\% instead of 50\%.

The can be explained if we take a closer look at the \hpp{} crypto
algorithm structure. 
Think of the lattice as a chessboard. A single plain text bit
may influence either the black or the white cells. More precisely, if
complementing bit $K$ influences black cells then, complementing bit
$K+4$ will influence white cells. Using $r+1$ rounds instead of $r$
will invert black and white. 
The truth is that a bit inversion in the clear text does not produce a
bit inversion of every ciphered bit with a probability 0.25. On the
contrary, a bit inversion in the clear text produce a bit inversion in
the ciphered text with a probability 0.5, for \emph{half the bits}.

Figure~\ref{fig:strict_text} shows that strict avalanche property
\emph{is not} true for this {\sc hpp} crypto algorithm. Instead it is
true for half the cells. 

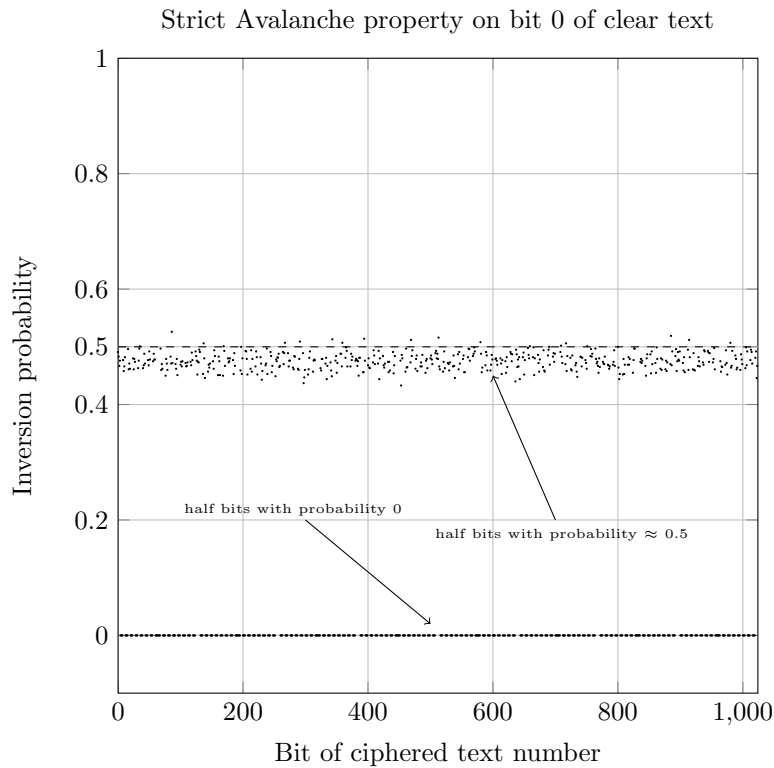
\begin{figure}[htbp]
\begin{center}
\begin{tikzpicture}
\begin{axis}[height=10cm, width=10cm,grid=major,
  xmin=0,xmax=1024,ymin=,ymax=1.0,
  extra y ticks={0.5},
  xlabel=Bit of ciphered text number,
  ylabel=Inversion probability,
  title=Strict Avalanche property on bit 0 of clear text]
\addplot[mark=none,draw=black,dashed] coordinates{(0,0.5) (1023,0.5)};
\addplot[only marks,mark=*,mark options={xscale=0.1,yscale=0.1}] file {data/strictavalanche_text_onebit.data};
\draw [->] (axis cs:300,0.2) -- (axis cs:500,0.02)
node[pos=-0.1]{\tiny half bits with probability 0};
\draw [->] (axis cs:700,0.2) -- (axis cs:600,0.45)
node[pos=-0.1]{\tiny half bits with probability $\approx$ 0.5};
\end{axis}
\end{tikzpicture}
\end{center}
\caption{Influence of changing a single bit in plain text :
only half of the bits is influenced
\label{fig:strict_text_one}}
\end{figure}

This can be emphasized if we reproduce the strict avalanche experiment
but only invert bit 0 (for instance) of the plain text. The result is
given on Figure~\ref{fig:strict_text_one}.

\section{Conclusion}
In this paper, we presented an original algorithm for symmetric
encryption based on the {\sc Hpp} lattice gas cellular automaton. 
The principle is to make data blocks evolve on a perturbed lattice.
The perturbations locations play the role of the encryption key.

It is proposed to use a large number of rounds, and to compress data,
before the encryption step. In any case, this algorithm should not be
used as such.

The fact that
{\sc Lgca} are used to simulate complex
dynamic systems~\cite{chop98,chopard06} lead us to think that there is no
obvious shortcut to compute the final configuration of the cryptographic
algorithm. Let us point out the pros and cons of {\sc hpp}
cryptography:
\begin{itemize}
\item Very large key space
\item Avalanche and strict avalanche properties are satisfied for key
  bits
\item Avalanche and strict avalanche properties are not satisfied for
  plain text bits. Nevertheless, the properties are satisfied for half
  the encrypted block.
\item {\sc hpp} cryptography is slow. But parallel computation of
  cellular automata is conceivable.
\item Size of encrypted blocks can be modified. Using large blocks
  increases size of the key space.
\end{itemize}

We plan to work on a modified (time reversible) \textsc{fhp} version of
this algorithm as using an hexagonal lattice is likely to solve the
"checker problem" and give better strict avalanche property results.
\bibliography{biblio_auto}

\begin{thebibliography}{10}
\expandafter\ifx\csname url\endcsname\relax
  \def\url#1{\texttt{#1}}\fi
\expandafter\ifx\csname urlprefix\endcsname\relax\def\urlprefix{URL }\fi
\expandafter\ifx\csname href\endcsname\relax
  \def\href#1#2{#2} \def\path#1{#1}\fi

\bibitem{hard73}
J.~Hardy, Y.~Pomeau, O.~de~Pazzis, Time evolution of a two-dimensional model
  system, J. Math. Phys 12 (1973) 1746--1759.

\bibitem{wolfram86}
S.~Wolfram, Random sequence generation by cellular automata, Adv. Appl. Math 7
  (1986) 123--169.

\bibitem{Seredynski03}
F.~Seredynski, P.~Bouvry, A.~Y. Zomaya, Cellular automata and symmetric key
  cryptography systems, in: Springer (Ed.), Genetic and Evolutionary
  Computation - Gecco, 2003, pp. 1369--1381.

\bibitem{tomassini00}
M.~Tomassini, M.~Parrenoud, Stream ciphers with one and two dimensionnal
  cellular automata, in: Springer (Ed.), Parallel Problem Solving from Nature -
  PPSN, 2000, pp. 722--731.

\bibitem{tomassini01}
M.~Tomassini, M.~Parrenoud, Cryptography with cellular automata, Applied Soft
  Computing (2001) 151--160.

\bibitem{wolfram85}
S.~Wolfram, Cryptographyy with cellular automata, in: Springer (Ed.), Advances
  in Cryptology : Crypto'85, 1985, pp. 428--432.

\bibitem{Seredynski2004753}
F.~Seredynski, P.~Bouvry, A.~Y. Zomaya, Cellular automata computations and
  secret key cryptography, Parallel Computing 30~(5-6) (2004) 753--766.

\bibitem{guan87}
P.~Guan, Cellular automaton public-key cryptosystem, Complex Systems 1 (1987)
  51--56.

\bibitem{kari92}
J.~Kari, Cryptosystems based on reversible cellular automata, Tech. rep.
  (1992).

\bibitem{gutowitz93}
H.~Gutowitz, Cryptography with dynamical systems.

\bibitem{seredynski04}
M.~Seredynski, P.~Bouvry, Block Encryption Using Reversible Cellular Automata,
  Vol. 3305, Springer, 2004, pp. 785--792.

\bibitem{chopard06}
S.~Marconi, B.~Chopard, Discrete physics, cellular automata and cryptography,
  in: 7th International Conference on Cellular Automata for Research and
  Industry, ACRI, Vol. 4173 of Lecture Notes in Computer Science, 2006, pp.
  617--626.

\bibitem{ulam52}
S.~M. Ulam, Random processes and transformations, in: Proceedings of the
  International Congress of Mathematicians, 1952.

\bibitem{neum66}
J.~V. Neumann, Theory of Self-Reproducing Automata, University of Illinois
  Press, 1966.

\bibitem{wolf00}
D.~A. Wolf-Gladrow, Lattice-Gas Cellular Automata and Lattice Boltzmann Models,
  Springer, 2000.

\bibitem{fris86}
U.~Frisch, B.~Hasslacher, Y.~Pomeau, Lattice-gasz automata for the
  navier-stockes equation, Phys. Rev. Lett. 14 (1986) 1505--1508.

\bibitem{feistel73}
H.~Feistel, Cryptogrpahy and computer privacy, scientific american 228~(5)
  (1973) 15--23.

\bibitem{webster85}
A.~Webster, S.~Tavares, On the design of s-boxes, in: Advances in cryptology :
  Crypto'85, Springer, 1985, pp. 429--432.

\bibitem{chop98}
B.~Chopard, M.~Droz, Cellular automata Modeling of Physical Systems, Cambridge
  University Press, 1998.

\end{thebibliography}
\end{document}